\documentclass[conference]{IEEEtran}
\usepackage{cite}
\usepackage{amsmath,amssymb,amsfonts,color,colortbl}
\usepackage[font=footnotesize]{caption}
\usepackage{algorithmic}
\usepackage{multirow,makecell,textgreek}
\usepackage{balance}
\usepackage{float,flushend}
\usepackage{graphicx}\usepackage{subcaption}
\usepackage[export]{adjustbox}
\def\BibTeX{{\rm B\kern-.05em{\sc i\kern-.025em b}\kern-.08em
    T\kern-.1667em\lower.7ex\hbox{E}\kern-.125emX}}

\begin{document}
\title{Reconfigurable Intelligent Surfaces for THz: Hardware Impairments and Switching Technologies}
\author{\IEEEauthorblockN{  
S\'{e}rgio Matos$^1$, Yihan Ma$^2$, Qi Luo$^2$, Jonas Deuermeier$^3$, Luca Lucci$^4$, Panagiotis Gavriilidis$^5$, \\Asal Kiazadeh$^3$, Verónica Lain-Rubio$^6$, Tung D. Phan$^7$, Ping Jack Soh$^7$, Antonio Clemente$^4$, \\Lu\'{i}s M. Pessoa$^8$, and George C. Alexandropoulos$^5$
}                   
\IEEEauthorblockA{$^1$University Institute of Lisbon, Portugal, $^2$University of Hertfordshire, UK, $^3$NOVA University Lisbon, Portugal, \\$^4$CEA-Leti, France, $^5$National and Kapodistrian University of Athens, Greece, $^6$ACST, Germany, \\$^7$University of Oulu, Finland, $^8$INESC TEC, Portugal}
}

\maketitle
\begin{abstract}
The demand for unprecedented performance in the upcoming 6G wireless networks is fomenting the research on THz communications empowered by Reconfigurable Inteligent Surfaces (RISs). A wide range of use cases have been proposed, most of them, assuming high-level RIS models that overlook some of the hardware impairments that this technology faces. The expectation is that the emergent reconfigurable THz technologies will eventually overcome its current limitations. This disassociation from the hardware may mask nonphysical assumptions, perceived as hardware limitations. In this paper, a top-down approach bounded by physical constraints is presented, distilling from system-level specifications, hardware requirements, and upper bounds for the RIS-aided system performance. We consider D-band indoor and outdoor scenarios where a more realistic assessment of the state-of-the-art solution can be made. The goal is to highlight the intricacies of the design procedure based on sound assumptions for the RIS performance. For a given signal range and angular coverage, we quantify the required RIS size, number of switching elements, and maximum achievable bandwidth and capacity. 
\end{abstract}

\begin{IEEEkeywords}
Reconfigurable intelligent surfaces, hardware impairments, beam squint, switches, D-band, unit cell design, use cases.
\end{IEEEkeywords}

\section{Introduction}
\begin{figure} [!t]
     \centering
     \includegraphics[width=\columnwidth]{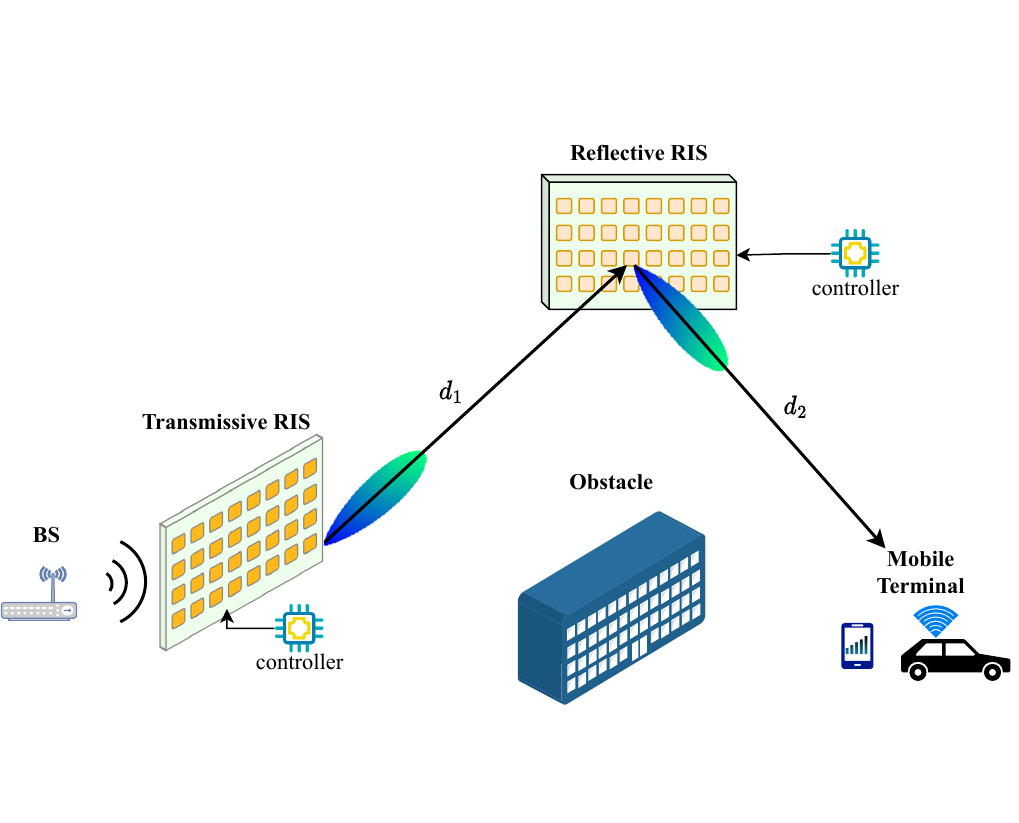}
    \caption{A typical wireless communications scenario includes two types of RISs: a transmissive RIS for efficient beamforming at the Base Station (BS) side and a reflective RIS to extend the coverage towards a mobile terminal. The reconfigurability of the RIS functionality is managed by a dedicated controller.}
     \label{fig:RIS-assisted}
\end{figure}

The Smart Networks and Services (SNS) Joint Undertaking (JU) is leading Europe's research and innovation towards pre-commercial 6G systems, expected in $2030$. Reconfigurable Intelligent Surfaces (RISs) have been envisioned as one of the main components that can support future high capacity smart radio environments~\cite{Tsinghua_RIS_Tutorial,RIS_challenges}. 
Several of the recently proposed 6G-related use cases are based on RIS-assisted wireless connectivity (a typical example is depicted in Fig.~\ref{fig:RIS-assisted}), differing on the type of terminals (e.g., factory robots, cars, and UAVs), type of coverage (indoor, outdoor, and outdoor-to-indoor), and type of RIS (e.g., transmissive or reflective RISs~\cite{RISoverview2023}). The goal is to extend the current capacity by progressing from millimeter-wave to THz wireless links. 

The development of cost-effective hardware solutions for operating at high frequencies (such as sub-THz and THz) is one of the main bottlenecks for the RIS technology. As the development of commercial THz reconfigurable hardware components is still in its infancy, the D-band (i.e., $110-170$\,GHz) provides a good playground for benchmarking different technologies. Moreover, D-band provides a good compromise between communication range and bandwidth; there has been intense research focused on this band~\cite{DBand_research}. The connection between high-level models of 6G communication systems with the RIS building blocks is intricate and difficult to quantify, as the THz technology is still evolving. The hiatus between the application and hardware strikes the ongoing research that sometimes oversimplifies the physical impairments of RISs. 

In this paper, we provide a more realistic link-budget calculation for two typical ranges of indoor and outdoor applications (20 and 100\,m) at $140$ GHz by considering key RIS hardware impairments. Namely, the maximum angular coverage that a RIS can provide, its implication on the beam direction as a function of frequency (beam squint~\cite{beamSquint}), quantization bits, and RIS insertion losses. The maximum bandwidth and corresponding capacity of the system are then estimated based on the required RIS size. Furthermore, the energy efficiency of the RIS is discussed considering some of the emergent reconfigurable technologies at D-band, such as phase-change materials (PCM)~\cite{Clemente24}, microfluidics~\cite{ yan2017adaptable}, memristors~\cite{Kim2024}, Gallium Nitride (GaN)~\cite{Lan23}, Bipolar Complementary Metal–Oxide–Semiconductor (BiCMOS)~\cite{Rebeiz23}, and Radio Frequency Silicon On Insulator (RF-SOI)~\cite{RF-SOI}. Finally, an example of a RIS unit-cell at D-band is provided to highlight some of the challenges related to the integration of the reconfigurable technology. We expect that this cross-domain description of RIS-assisted communications systems can foment closer collaboration between the hardware and the communications/signal processing communities. 

The paper is organized as follows. In Section~II, the considered hardware impairments are described, namely maximum angular coverage (II.A), RIS phase quantization (II.B), bandwidth limitations (II.C) and aperture efficiency (II.D). In Section~III, link budget analysis is performed based on the defined indoor and outdoor scenarios, while Section~IV focuses on the implementation of the RIS unit cell. Different reconfigurable technologies are benchmarked in terms of integration efficiency, switching performance, and power efficiency. The main conclusions of the paper are drawn in Section V.

\section{RIS Hardware Impairments}
The most common core components of a RIS are~\cite{Tsinghua_RIS_Tutorial,RISoverview2023}: \textit{i}) Radio-Frequency (RF) switches, handling the reconfigurability of its elements; \textit{ii}) unit cells that convert the Direct Current (DC) switch actuation into RF controlled signals; and \textit{iii}) panel hosting the unit cells which act as a reconfigurable planar antenna array/metasurface enabling dynamic control of the radio environment for communications and/or sensing~\cite{HRIS_Mag}. Each of these components imposes certain limitations to the RIS performance, as will be described in the sequel.

\subsection{Angular Coverage}
\begin{figure} [!t]
     \centering
     \includegraphics[width=\columnwidth]{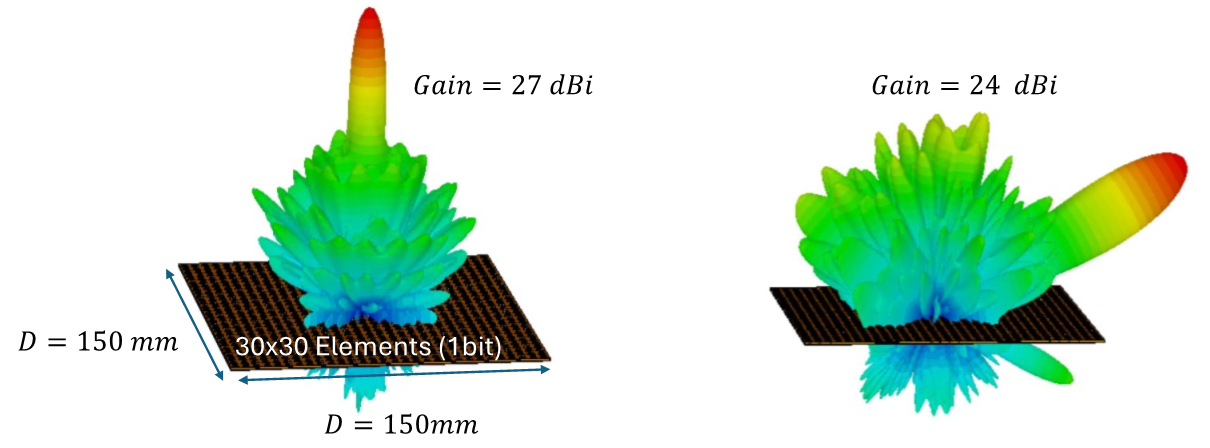}
    \caption{Full-wave evaluation of a RIS composed by $30\times30$ unit cells ($300\times300~\text{mm}^{2}$) operating at $30$~GHz with $1$-bit phase reconfigurability (PIN-diode) when illuminated by a near-field source.}
     \label{fig:RIS_angle_coverage_30_GHz}
\end{figure}

The required beam steering angular range of the RIS, \(\theta_{\max}\), varies with the use case. For example, for outdoor use cases, it is highly desirable to have this range as large as possible to provide a larger coverage area. However, a planar metasurface is bounded by the following physical constraint: its effective aperture decreases with the cosine of the observation angle. The angular coverage up to \(\theta_{\max} = 60^{\circ}\) limits these scanning losses to $3$~dB, which seems as a good figure to account for in the system design. As will be shown later, this angle also influences how the beam pointing direction varies with frequency (beam squint), hence, it should not be taken lightly. Instead, a RIS needs to be adjusted to the specific use case but considering the aforementioned limit. In Fig.~\ref{fig:RIS_angle_coverage_30_GHz}, we present an example full-wave simulation of a complete RIS model with $30\times 30$ unit cells, each with $1$-bit phase reconfigurability realized via a PIN-diode, designed for $30$~GHz {\cite{Eucap2024KaRISPin}}, which confirms the expected $3$ dB scan loss caused by the $60^{\circ}$ beam tilt.

\subsection{Response Quantization}
\begin{figure} [!t]
     \centering
     \includegraphics[width=\columnwidth]{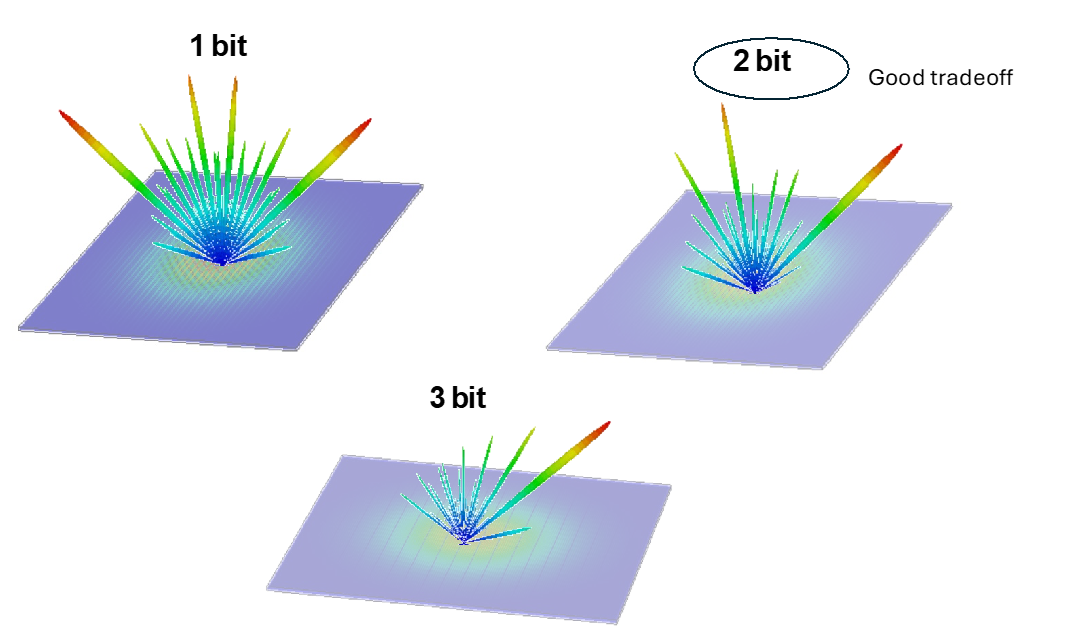}
    \caption{Numerical evaluation of an idealized RIS aperture with diameter $111$~mm operating at $140$~GHz that emulates an outgoing tilted plane wave (\(45^{\circ})\) generated with different phase quantization steps (1/2/3 bits).}  \label{fig:RIS_phase_quantization}
\end{figure}

The switch-based reconfigurability of the RIS implies that the phase control of the metasurface is quantized. To implement more phase states, a higher number of switches per unit cell needs to be integrated. Hence, a balance needs to be achieved between RIS performance and switch integration complexity. Figure~\ref{fig:RIS_phase_quantization} includes a full-wave numerical evaluation of the phase quantization effect for RIS reflective beamforming for the case where a RIS is designed to reflect an incident normal plane wave into an outgoing plane tilted at \(45^{\circ}\), e.g., for the signal coverage example in Fig.~\ref{fig:RIS-assisted}. It can be concluded that a $2$~bit quantization offers a good comprise. Another factor to take into account is that the number of switches required per bit is not necessarily one. As discussed in {\cite{Eucap2024KaRISPin}}, improved unit-cell performance, in terms of bandwidth, can be obtained when a bit is generated by inter-commuting two switches simultaneously in opposite states. Moreover, when independent polarization control is required the number of switches doubles. Considering these facts, it seems more realistic to consider that $4$ switches per unit cell will be a typical requirement for many of the defined use cases.

\subsection{Bandwidth and Beam Squint}
The RIS operation bandwidth depends on multiple factors~\cite{RIS_challenges}. Apart from the obvious dependence on the unit-cell bandwidth, it also depends on the dispersion introduced by the collective effect of the elements that compose the aperture. Even with ideal unit cells and switch devices, the aperture will exhibit a dependence on the beam-pointing direction with frequency (beam squint). This effect is well-known from array theory and becomes relevant when operating with very narrow beams, as is the case with large apertures. The received power will decrease with frequency as the beam shifts. A $3$~dB bandwidth can be associated with this intrinsic physical effect, defining an upper bound for the maximum bandwidth that a given RIS can achieve. 

\begin{figure} [!t]
     \centering
     \includegraphics[width=\columnwidth]{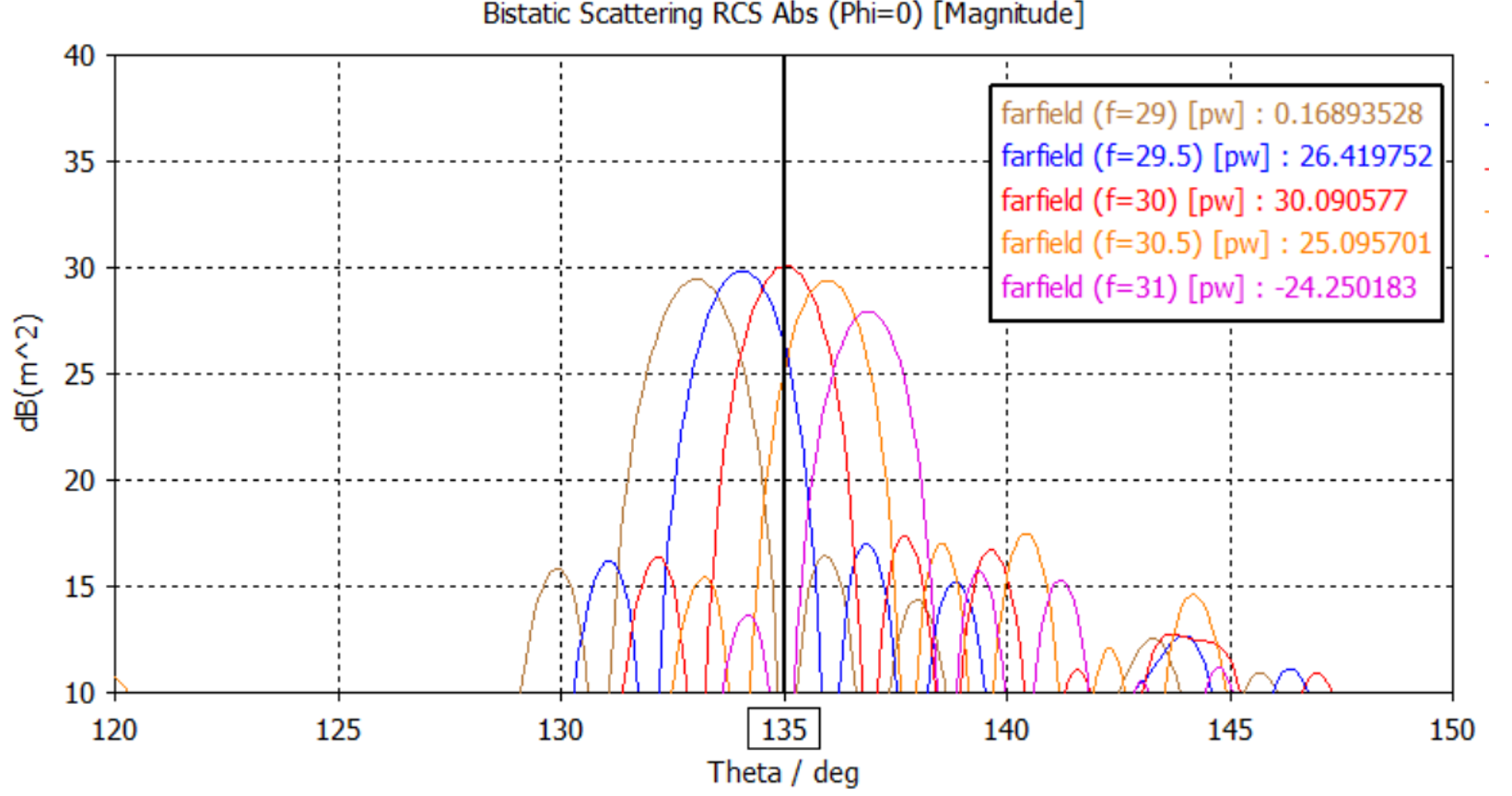}
    \caption{Radar cross-section of a $30\times30$ RIS with $1$-bit phase resolution at $30$ GHz designed to reflect and tilt a normal incident plane wave to \(45^{\circ}\) (this corresponds to \(135^{\circ}\) in the adopted frame for the simulation model).}
     \label{fig:RIS_RCS_30_GHz}
\end{figure}

In Fig.~\ref{fig:RIS_RCS_30_GHz}, considering the designed RIS in Fig.~\ref{fig:RIS_angle_coverage_30_GHz}, we depict how the beam squint limits the bandwidth of the system. It can be seen that, for a given observation angle of the moving terminal (in this case \(135^{\circ}\)), the intensity of the received power depends on the operating frequency, with the maximum taking place only at the designed frequency (in this case \(30\) GHz).
This shift with frequency depends both on the dimension of the RIS and on the observation angle. The higher each of these parameters is, the more beam shift will occur, and a smaller bandwidth will be available. This less obvious relation between angular coverage and maximum system bandwidth is quantified in Tables ~\ref{tab:outdoor_requirements} and ~\ref{tab:indoor_requirements}.

\subsection{Aperture Efficiency}
The aperture efficiency provides crucial information on how efficient is the physical area of the RIS panel for generating a reflective beam. A \(100~\%\) aperture efficiency indicates that the RIS is as efficient as a metallic plate for reflecting a normal incident plane wave. However, this ideal value is never achieved with metasurfaces, since the RIS aperture efficiency depends in practice on many factors, including the phase quantization error and the unit-cell insertion losses. For passive RIS structures, the literature reports that more than \(50~\%\) aperture efficiency can be hardly achieved~\cite{AppertureEfficiency}. In addition, for RISs, we have the additional complexity of integrating the reconfigurable elements and the corresponding bias circuitry. Based on previous works on reconfigurable technologies for RIS unit cells~\cite{insertion_loss_table}, at least $3$~dB of insertion losses are expected. Therefore, \(25~\%\) aperture efficiency can be considered as a realistic figure for defining an efficient RIS in terms of RF performance.

\section{RIS-Aided-Link Budget Analysis}\label{sec:Link}
In this section, we consider two representative scenarios compatible with indoor and outdoor use cases at $140$~GHz, whose parameters are summarized in Table~\ref{tab:specifications}, to evaluate the performance of RIS-aided wireless links when the previously RIS hardware impairments are taken into account. For this assessment, we have made the following assumptions: 
\begin{itemize}
\item  To overcome free-space path loss, it was assumed that the combined gain of the antennas at the terminal and the BS equals to \(56\)~dBi, which is compatible with typical scenarios of millimeter-wave links. 
\item The available generated power at the sub-THz frequency of $140$ GHz using the technology in \cite{8874296} leads to a modulated power radiated by the BS of \(20\)~dBm. 
\item The noise can be estimated based on the receiver terminal's noise figure. For the considered scenarios, the noise figure was set to \(5\)~dB and the noise power density to \(-174\) ~dBm/Hz~\cite{SUYAMA20212020FGI0002}. 
\item The minimum received power is such that, for a reference \(10\)~GHz of bandwidth, the Signal-to-Noise (SNR) is \(10\) ~dB, enabling high capacity wireless communications following the theoretical limit given by the Hartley-Shannon theorem.
\end{itemize}

The required size of the RIS panel can be obtained from a link budget analysis based on a bi-static radar formulation, as recently shown in~\cite{Terraameta_EuCAP2024}. Having defined the RIS size, the number of required unit cells included in the panel can be calculated, assuming the typical value of \(\lambda/2\) for inter-cell spacing~\cite{RIS_rich_scattering}, with $\lambda$ being the free-space wavelength. After defining the phase quantization and the number of switches for implementing the per bit quantization, as previously explained, the total number of switches required for the RIS aperture can be computed, which provides a good picture of the underlying complexity required for each RIS unit cell. This analysis also provides the maximum capacity calculated, via the Shannon–Hartley theorem, using SNR value and the maximum $3$~dB bandwidth of the system due to beam squint (as described previously). 

The RIS's requirements for the two considered scenarios at $140$ GHz are summarized in Tables~\ref{tab:outdoor_requirements} and \ref{tab:indoor_requirements}. It is clearly shown that, as the size of the RIS increases, the available maximum bandwidth is significantly reduced, with this reduction being larger when larger coverage angles are required. It is important to highlight that this result is not influenced by the bandwidth of the unit cell. This implies that, even if ultra-wideband THz unit cells were available, these system restrictions would still hold. Therefore, these factors need to be carefully accounted when designing RIS-aided wireless systems with large RIS apertures. This problem also happens as we increase the operating frequency towards THz, which requires compensating the increasing free-space path losses~\cite{Transmitarray_THz_2023}. Hence, for a given use case, it is necessary to optimize the RIS angular coverage, the link distance, and operation frequency. Otherwise, as we move towards THz, the RIS-aided wireless system will yield lower effective capacity.  

\begin{table}[!t]
\centering
\caption{\centering System specifications for representative indoor and outdoor scenarios at $140$ GHz where the wireless link is established via a reflective RIS (see Fig.~\ref{fig:RIS-assisted} for \(d_1=d_2\), thus, the table includes range values for the length \(d_1+d_2\) in meters).}
\begin{tabular}{|m{4cm}|m{1.5cm}|m{1.5cm}|}
\hline
\multirow{2}{*}{\textbf{Specification}} & \multicolumn{2}{c|}{\textbf{Scenarios}} \\ \cline{2-3} 
                                & \multicolumn{1}{>{\centering\arraybackslash}m{1.5 cm}|}{\textbf{Outdoor }} & \multicolumn{1}{>{\centering\arraybackslash}m{1.5 cm}|}{\textbf{Indoor}}   \\ \hline
Range (meters)                  &  \multicolumn{1}{>{\centering\arraybackslash}c|}{\(100\)}               &  \multicolumn{1}{>{\centering\arraybackslash}c|}{\(20\)}                      \\ \hline
Total antenna gain (dBi)     & \multicolumn{2}{>{\centering\arraybackslash}c|}{\(56\)}       \\ \hline
Radiated power (dBm)             & \multicolumn{2}{>{\centering\arraybackslash}c|}{\(20\)}         \\ \hline
Noise power density (dBm/Hz)         & \multicolumn{2}{>{\centering\arraybackslash}c|}{\(-174\)}  \\ \hline
Receiver noise figure (dB)         & \multicolumn{2}{>{\centering\arraybackslash}c|}{\( 5\)}            \\ \hline
Reveived power (dBm)            & \multicolumn{2}{>{\centering\arraybackslash}c|}{\( -59\) }      \\ \hline
Phase quantization (\# bits)          & \multicolumn{2}{>{\centering\arraybackslash}c|}{\(2\)}          \\ \hline
\end{tabular}
\label{tab:specifications}
\end{table}

\begin{table}[!t]
\centering
\caption{\centering System performance and RIS's requirements for the outdoor scenario in Table~\ref{tab:specifications} at $140$ GHz.}
\begin{tabular}{|m{2cm}|m{2.5cm}|m{2.5cm}|}
\hline
\multicolumn{3}{|>{\centering\arraybackslash}c|}{\textbf{Outdoor Scenario}}                      \\ \hline
\multicolumn{3}{|>{\centering\arraybackslash}c|}{\textbf{RIS requirements}}                      \\ \hline
                                    &  \(\theta_{\max}= 50^{\circ}\) & \(\theta_{\max}= 60^{\circ}\)      \\ \hline
RIS size (mm$^2$)             &     \(118 \times 118\)         & \(125 \times 125\)       \\ \hline
\# unit cells                       &      \(111 \times 111\)        &   \(118 \times 118\)       \\ \hline
\# switches \((\times 10^3)\) (\(2\) switches per bit) & \(48.56\)   & \(55.06\)          \\ \hline
\multicolumn{3}{|>{\centering\arraybackslash}c|}{\textbf{System performance}}                      \\ \hline
Maximum \(3\)~dB bandwidth (GHz)     & \(2.4\, (1.7 \%)\)             & \(1.5\, (1.1\%)\)     \\ \hline
SNR (dB)                            &        16                      &     18     \\ \hline
Achievable capacity (Gbps)          &       13                       &     9.3     \\ \hline
\end{tabular}
\label{tab:outdoor_requirements}
\end{table}

\begin{table}[!t]
\centering
\caption{\centering System performance and RIS's requirements for the indoor scenario in Table~\ref{tab:specifications} at $140$ GHz.}
\begin{tabular}{|m{2cm}|m{2.5cm}|m{2.5cm}|}
\hline
\multicolumn{3}{|>{\centering\arraybackslash}c|}{\textbf{Indoor Scenario}}                      \\ \hline
\multicolumn{3}{|>{\centering\arraybackslash}c|}{\textbf{RIS requirements}}                      \\ \hline
                                    &  \(\theta_{\max}= 50^{\circ}\) & \(\theta_{\max}= 60^{\circ}\)      \\ \hline
RIS size (mm$^2$)             &     \(24 \times 24\)         & \(26 \times 26\)       \\ \hline
\# unit cells                       &      \(22 \times 22\)        &   \(24 \times 24\)       \\ \hline
\# switches \((\times 10^3)\) (\(2\) switches per bit) & \(1.94\)   & \(2.20\)          \\ \hline
\multicolumn{3}{|>{\centering\arraybackslash}c|}{\textbf{System performance}}                      \\ \hline
Maximum \(3\)~dB bandwidth (GHz)     & \(12\, (8.6 \%)\)             & \(7.8\, (5.6\%)\)     \\ \hline
SNR (dB)                            &        9.2                      &     11.1     \\ \hline
Maximum capacity (Gbps)          &       38                       &     29     \\ \hline
\end{tabular}
\label{tab:indoor_requirements}
\end{table}

\section{RIS Unit-cell Design at High Frequencies}
\subsection {D-band Unit-Cell and RIS Example}
\begin{figure}[!t]
    \centering
    \includegraphics[width=7cm]{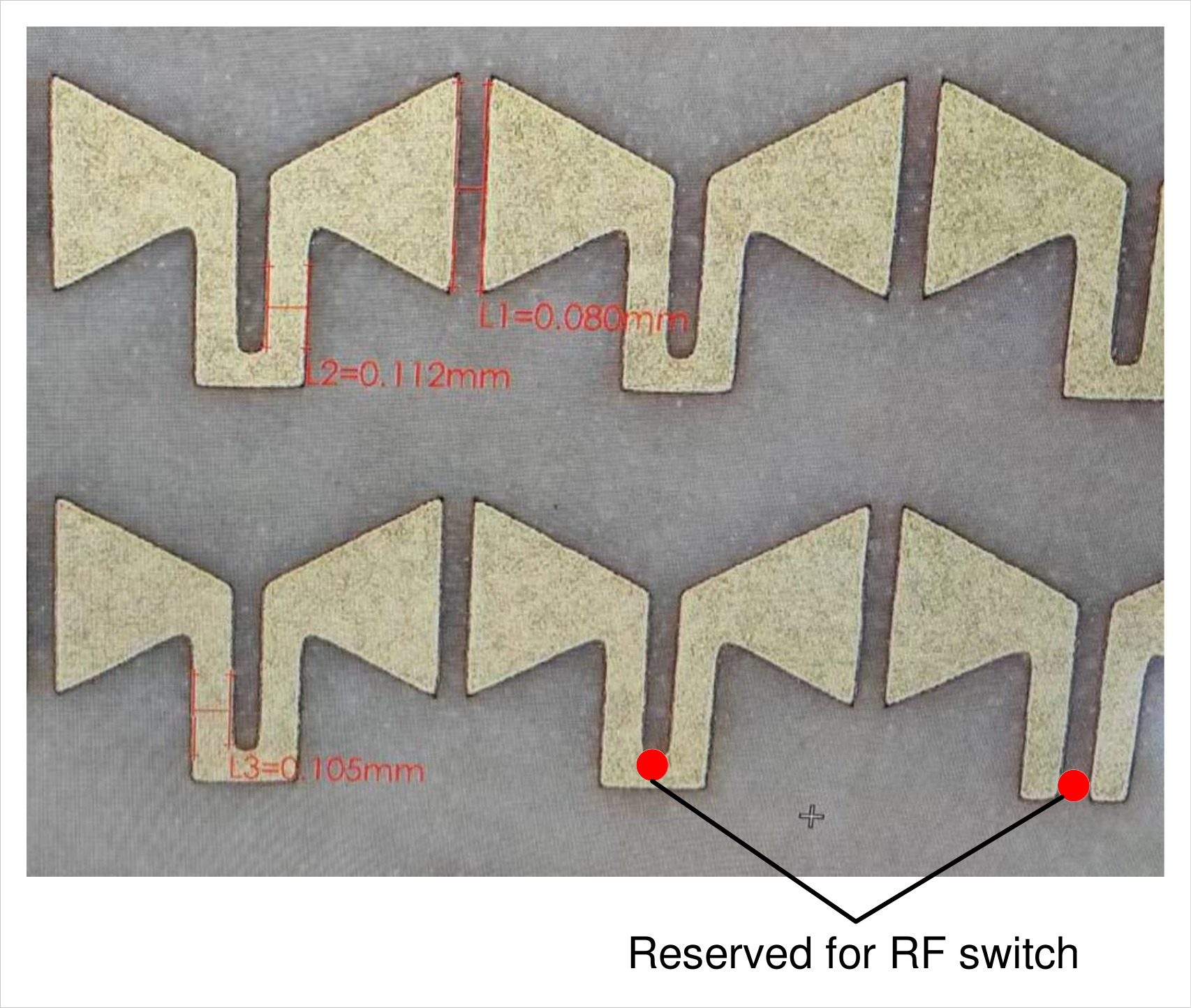}
    \caption{Photo of the fabricated \(1\)-bit RIS prototype at sub-THz in~\cite{Eucap2024RISfab}.}
    \label{fig:TCDA_photo_1}
\end{figure}

\begin{figure}[!t]
\centering
\includegraphics[width=\columnwidth]{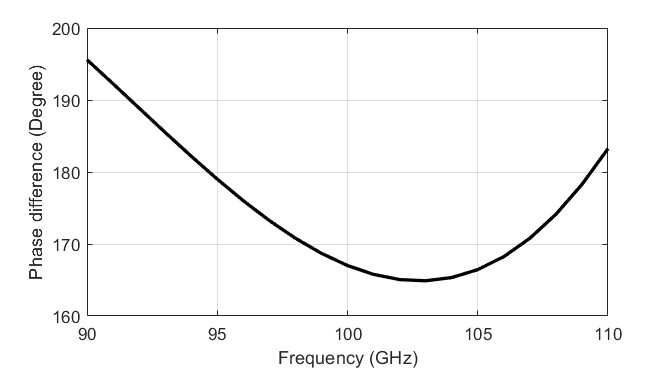}
\caption{The reflection phase difference of the designed $1$-bit RIS unit cell when the RF switch is ``ON'' and ``OFF'' as a function of the operating frequency.}
\label{fig:figure_TCDA_phase}
\end{figure}

An example of an \(1\)-bit RIS unit cell at D-band is provided in this section to better describe the associated integration challenges relating to the unit-cell RF design and the switching technology. A passive RIS prototype is illustrated in Fig.~\ref{fig:TCDA_photo_1} that mimics the use of a $1$-bit phase quantization step~\cite{Eucap2024RISfab}. Due to the lack of available RF switches at the sub-THz frequency band, a microstrip was used to mimic the case when the RF switch is ``ON'' and an open circuit was used to mimic the case when the RF switch is ``OFF.'' This RIS was designed to operate at the center frequency of $102$ GHz. The dimension of the meta-atom is \(1.2 \times 1.2\)~mm. It can be seen from Fig.~\ref{fig:TCDA_photo_1} that, to fit this design, the future RF switch to be integrated needs to have a length of no more than \(80\)~\textmu m, and the length of the switch with its package should be no more than \(280\)~\textmu m. This already restricts the use of some of the commercially available  RF PiN diode-based solutions. The phase response of the RIS shown in Fig.~\ref{fig:TCDA_photo_1} would be achievable by changing the phase of the reflected wave by 180$^\circ$ through the control of an RF switch. Figure~\ref{fig:figure_TCDA_phase} shows the simulated reflection phase differences of the designed RIS for the two states of an RF switch. It can be seen that the unit cell exhibits a wideband response with a $180^o\pm20^o$ phase difference from the frequency $90$ to $110$ GHz for the two different states of the RF switch.
 
\subsection{Overview of RF Switch Technologies}
The design of RISs in the D-band and beyond poses several major challenges on the choice of the switch element. The ideal switch should be of high performance, i.e., featuring low insertion losses and good isolation, which corresponds in switch-jargon to a very low $\textrm{R}_\textrm{on}\times\textrm{C}_\textrm{off}$ figure of merit. Equivalently, in some scientific communities, a very high cut-off frequency, $F_C\triangleq1/(2\pi\,\textrm{R}_\textrm{on}\textrm{C}_\textrm{off})$, is an equivalent figure-of-merit. In addition, the ideal switch should also be as small as possible. See, for example, Fig.~\ref{fig:TCDA_photo_1} where the required size is no more than 80~$\mu$m in the largest lateral dimension, although, the smaller the better. Moreover, the switch should have no measurable or negligible DC dissipation when idle and commute between the ``ON'' and ``OFF'' states with a minimal amount of dissipated energy. Last but not least, the switch should be easily integrable in close proximity with the antenna (luckily, this can be achieved with most technologies) but also be integrable on the same substrate as the digital and analog circuitry needed to drive the switch itself, to limit off-chip communications to digital and clock signals. The last requirement is one that most research-level technologies have yet to be proven upon. 

In Table~\ref{tab:switch-tech}, an overview of the aforementioned requirements is presented together with, whenever possible, a quantitative assessment for each of the technologies considered. Having limited resources, the choice has been limited to three commercial- and three research-level technologies: the distinction is reflected in the Technology Readiness Level (TRL) line. For each technology, to the best of the authors' knowledge, the highest working frequency of the switch is reported, as found in the literature. At the current state, there is no clearly indisputable winner and the best choice for a D-band RIS necessitates further research.

\begin{table}[!t]
\centering
\caption{\centering Comparison of switching technologies for sub-{TH}z RISs using data collected from~\cite{Sobo22,Liu22,Mansur24,Wain21}, supplemental material of~\cite{Wain21}, and references therein. Switching energy is 
listed only when DC dissipation is negligible.}
\begin{tabular}{|m{1.5cm}|m{0.7cm}|m{0.7cm}|m{0.7cm}|m{0.7cm}|m{0.7cm}|m{0.7cm}|}
\hline

& \multicolumn{6}{|>{\centering\arraybackslash}m{6cm}|}{\textbf{Reconfigurable Technology Selection for THz}}  \\ \hline
& \rotatebox{90}{RF-SOI} & \rotatebox{90}{BiCMOS} & \rotatebox{90}{GaN-on-Si} & \rotatebox{90}{Microfluidics\,
} & \rotatebox{90}{Memristors\,} & \rotatebox{90}{PCM}  \\ \hline 
Max freq. in\newline publ. [GHz]     & 220  & 133  & 40   & 123   & 480 & 67   \\ \hline
Switch size                  & ++  & ++  & +    & +   & +++  & +++   \\ \hline
CMOS \newline integration & +++  & +++  & ++   & +   & +   & ++   \\ \hline
$\textrm{R}_\textrm{on}\times\textrm{C}_\textrm{off}$ [fs] &  90  &   80   &   55 &  TBD & $<$10 & $<$10 \\ \hline
DC dissipation \newline [mW]                &  0.05-0.1    &   10-50     &  0.1-1     &    0.001    &  none    &  none  \\ \hline            
Switching \newline energy [nJ]      &      &        &      &     20   &   1-10  &  1-500  \\ \hline            
TRL                       &   9  & 8    & 6    & 1-3 & 2  & 4  \\ \hline
References                     & {\tiny\hspace{-0.7em}~\cite{Rebeiz12,Seo23}} & {\tiny\hspace{-0.8em}~\cite{Cress14,gemic18}}  &    
   {\tiny\hspace{-0.8em}~\cite{Intel20,Longhi24} }  & {{\tiny\hspace{-0.7em}~\cite{reichel2018electrically,jackel1982electrowetting}}}    &   {\tiny\hspace{-0.7em}~\cite{Kim2024}}  &  {\tiny{\hspace{-1em}}~\cite{Kim2024,Mansur24}}    \\ \hline
\end{tabular}
\label{tab:switch-tech}
\end{table}

\subsubsection{Conventional Technologies}
The large quantity of switching elements included in the RISs considered for the scenarios in Section~\ref{sec:Link} indicate the underlying complexity required for the RIS design with commercial or conventional technologies. In addition, a fundamental aspect of the 6G specifications is the strict requirement for power efficiency. It is then important to quantify the DC power consumption required to operate these RISs. Depending on the technology of choice, the DC dissipation of the switch element can already sum up to huge numbers, especially for common commercial technologies. With PiN or Heterojunction Bipolar Transistor (HBT) switches based on Silicon Germanium (SiGe), DC consumption can reach 10s~mW  which, for the outdoor scenario, could represent an unacceptable kW level of dissipated power just to operate the switching part of the RIS. 

As an example, the exact DC consumption for a BiCMOS PiN-diode-based switch can be found using~\cite{Cress14}. Namely, the DC power consumption of the Single Pole, Double Throw (SPDT) switch with one arm in the ``ON'' state is \(10.2\)~mW at \(1.2\)~V. The SPDT achieves less than \(2\)~dB insertion loss and good isolation between \(73\)-\(133\)~GHz. The corresponding integrated circuit occupies an area of \(580 \times 240\)~\textmu m$^2$ without RF pads, but including matching structures. Hence, the potential for this solution is quite interesting: a well-known diode-based approach demonstrated up to \(133\)~GHz and easily integrable with CMOS circuitry. However, this solution is hardly optimal from a power consumption perspective limiting its application when energy efficiency is one of the core design objectives. 

Other commercial solutions based on GaN High Electron Mobility Transistors (HEMTs) or RF-SOI provide much reduced power consumption compared to PiN diodes. RF-SOI switches are widely used in industry and have been demonstrated
up to \(220\)~GHz with excellent isolation and insertion losses\cite{Rebeiz12}. GaN-on-Silicon is also a promising solution: an attractive  $\textrm{R}_\textrm{on}\times\textrm{C}_\textrm{off}=55$~fsec has been recently measured~\cite{Intel20}, and the first switches working up to \(40\)~GHz  reported~\cite{Longhi24}. But GaN-on-Silicon still falls short of the expectation set by the much more expensive (and non-CMOS compatible) GaN-on-Silicon-Carbide, where SPST switches up to \(330\) GHz have been demonstrated\cite{Ambacher19}.

\subsubsection{Emerging Technologies}
Unlike conventional RF switches, non-volatile RF switches can operate without any DC static power dissipation, because of their non-volatile resistive switching effect, and are usually an order of magnitude smaller in occupied area. In this regard, disruptive new technologies are being developed, including memristors~\cite{Kim2024,Kiazadeh2023} and Phase-Change Materials (PCM) \cite{Clemente24}. Another emerging technology is microfluidics using electrical actuation \cite{reichel2018electrically}.

PCM-based RF switches based on Germanium Telluride (GeTe) have been demonstrated up to $67$~GHz with excellent isolation/insertion losses and a 
$15\times$ reduction in chip area compared to RF-SOI~\cite{Mansur24}. Even more promising are memristors: a monolayer molybdenum disulfide (MoS$_2$) switch was shown to exhibit excellent insertion losses and isolation at frequencies up to \(480\)~GHz with $\textrm{R}_\textrm{on}\times\textrm{C}_\textrm{off}=2.3$~fsec, which is lower than any other emerging RF switch technology~\cite{Kim2024} while consuming just $49$~pJ in the switching transient, which is the only thing that consumes power in the memristor. However, these promising findings need to be balanced with the fact that the integration of these new technologies still face numerous challenges before reaching a stage of industrial mass production, which will need a long development time.  

Another emergent reconfigurable technology is microfluidics which has several attractive properties when compared to semiconductor switches: a very high ``ON''–``OFF'' impedance ratio with ``OFF''-state impedance on the order of $10^{10}$– $10^{14}$~$\Omega$ \cite{sen2008microscale}, and low contact resistance at ``ON''-state (below $50$~m$\Omega$ as shown in~\cite{sen2008microscale}). The power consumption of a microfluidics-based switch is very much dependent on the physical dimensions of the switch structure, actuation method, and microfluidic system design. An example of such a switch using the Continuous ElectroWetting (CEW) actuating method was reported in~\cite{jackel1982electrowetting} with an operating voltage of around $1$~V and low power consumption around $1$~μW. However, the reported switching time was $20$~ms which is slower compared to those of semiconductor-based switches, as shown in Table~\ref{tab:switch-tech}. This fact can limit the application of this technology in some scenarios necessitating fast switching operations. Finally, the total switching energy of such a microfluidics-based switch is only $20$ nJ, a fact that highlights the possibility of designing highly energy efficient microfluidics-based RISs.

\section{Conclusion}
In this paper, we presented an encompassing study starting from conventional indoor and outdoor RIS-assisted communication scenarios at high frequencies. From these specifications, specific system performance indicators were correlated with the necessary hardware requirements, including a benchmark of different reconfigurable technologies. The goal of this study is to highlight the need for accounting hardware and physical impairments inherent to the RIS tehcnology. Hopefully, this discussion can contribute to bringing closer the hardware and the communications/signal processing research communities for more realistic assessments of the RIS technology.  

\section*{Acknowledgment}
This work was supported by the SNS JU project TERRAMETA under the European Union's Horizon Europe research and innovation programme under Grant Agreement No 101097101, including top-up funding by UK Research and Innovation (UKRI) under the UK government’s Horizon Europe funding guarantee. 

\bibliographystyle{IEEEtran}
\bibliography{IEEEabrv,references}

\begin{thebibliography}{10}
\providecommand{\url}[1]{#1}
\csname url@samestyle\endcsname
\providecommand{\newblock}{\relax}
\providecommand{\bibinfo}[2]{#2}
\providecommand{\BIBentrySTDinterwordspacing}{\spaceskip=0pt\relax}
\providecommand{\BIBentryALTinterwordstretchfactor}{4}
\providecommand{\BIBentryALTinterwordspacing}{\spaceskip=\fontdimen2\font plus
\BIBentryALTinterwordstretchfactor\fontdimen3\font minus
  \fontdimen4\font\relax}
\providecommand{\BIBforeignlanguage}[2]{{%
\expandafter\ifx\csname l@#1\endcsname\relax
\typeout{** WARNING: IEEEtran.bst: No hyphenation pattern has been}%
\typeout{** loaded for the language `#1'. Using the pattern for}%
\typeout{** the default language instead.}%
\else
\language=\csname l@#1\endcsname
\fi
#2}}
\providecommand{\BIBdecl}{\relax}
\BIBdecl

\bibitem{Tsinghua_RIS_Tutorial}
M.~Jian, G.~C. Alexandropoulos, E.~Basar, C.~Huang, R.~Liu, Y.~Liu, and
  C.~Yuen, ``Reconfigurable intelligent surfaces for wireless communications:
  Overview of hardware designs, channel models, and estimation techniques,''
  \emph{Intell. Converged Netw.}, vol.~3, no.~1, pp. 1--32, 2022.

\bibitem{RIS_challenges}
G.~C. Alexandropoulos, M.~Crozzoli, D.-T. Phan-Huy, K.~D. Katsanos,
  H.~Wymeersch, P.~Popovski, P.~Ratajczak, Y.~B{\'e}n{\'e}dic, M.-H. Hamon,
  S.~Herraiz~Gonzalez, R.~D'Errico, and E.~Calvanese~Strinati, ``{RIS}-enabled
  smart wireless environments: {D}eployment scenarios, network architecture,
  bandwidth and area of influence,'' \emph{EURASIP J. Wireless Commun. Netw.},
  vol. 103, pp. 1--38, 2023.

\bibitem{RISoverview2023}
E.~Basar, G.~C. Alexandropoulos, Y.~Liu, Q.~Wu, S.~Jin, C.~Yuen, O.~Dobre, and
  R.~Schober, ``Reconfigurable intelligent surfaces for {6G}: Emerging
  applications and open challenges,'' \emph{arXiv preprint arXiv:2312.16874},
  2023.

\bibitem{DBand_research}
T.~Maiwald \emph{et~al.}, ``A review of integrated systems and components for
  6g wireless communication in the {D}-band,'' \emph{Proc. IEEE}, vol. 111,
  no.~3, pp. 220--256, 2023.

\bibitem{beamSquint}
S.~Targonski and D.~Pozar, ``Minimization of beam squint in microstrip
  reflectarrays using an offset feed,'' in \emph{Proc. IEEE Antennas Prop. Soc.
  Int. Symp.}, vol.~2, 1996, pp. 1326--1329.

\bibitem{Clemente24}
S.~Gharbieh, J.~Milbrandt, B.~Reig, D.~Mercier, M.~Allain, and A.~Clemente,
  ``Design of a binary programmable transmitarray based on phase change
  material for beam steering applications in {D}-band,'' \emph{Sci. Rep.},
  vol.~14, no. 2966, 2024.

\bibitem{yan2017adaptable}
L.~B. Yan \emph{et~al.}, ``Adaptable metasurface for dynamic anomalous
  reflection,'' \emph{Appl. Physics Lett.}, vol. 110, no.~20, 2017.

\bibitem{Kim2024}
D.~Kim, S.~J. Yang, N.~Wainstein, S.~Skrzypczak, G.~Ducournau, E.~Pallecchi,
  H.~Happy, E.~Yalon, M.~Kim, and D.~Akinwande, ``Emerging memory electronics
  for non-volatile radiofrequency switching technologies,'' \emph{Nature Rev.
  Electr. Eng.}, vol.~1, pp. 10--23, 2024.

\bibitem{Lan23}
F.~Lan, L.~Wang, H.~Zeng, S.~Liang, T.~Song, W.~Liu, P.~Mazumder, Z.~Yang,
  Y.~Zhang, and D.~M. Mittleman, ``Real-time programmable metasurface for {THz}
  multifunctional wave front engineering,'' \emph{Light: Sci. Appl.}, vol.~12,
  no.~1, 2023.

\bibitem{Rebeiz23}
S.~Wang and G.~M. Rebeiz, ``Dual-band 28- and 39-{GHz} phased arrays for
  multistandard {5G} applications,'' \emph{IEEE Trans. Microw. Theory Tech.},
  vol.~71, no.~1, pp. 339--349, 2023.

\bibitem{RF-SOI}
C.~Wang and R.~Han, ``17.6 rapid and energy-efficient molecular sensing using
  dual mm-wave combs in 65nm cmos: A 220-to-320ghz spectrometer with 5.2{mW}
  radiated power and 14.6-to-19.5{dB} noise figure,'' in \emph{Proc. IEEE
  ISSCC}, San Francisco, USA, 2017.

\bibitem{HRIS_Mag}
G.~C. Alexandropoulos, N.~Shlezinger, I.~Alamzadeh, M.~F. Imani, H.~Zhang, and
  Y.~C. Eldar, ``Hybrid reconfigurable intelligent metasurfaces: {E}nabling
  simultaneous tunable reflections and sensing for {6G} wireless
  communications,'' \emph{IEEE Veh. Technol. Mag.}, to appear, 2024.

\bibitem{Eucap2024KaRISPin}
F.~Cardoso, S.~Matos, L.~M. Pessoa, A.~Clemente, J.~R. Costa, C.~A. Fernandes,
  and J.~M. Felício, ``Improved performance of a $1$-bit {RIS} by using two
  switches per bit implementation,'' in \emph{Proc. EuCAP}, Glasgow, Scotland,
  2024.

\bibitem{AppertureEfficiency}
M.~H. Dahri, M.~H. Jamaluddin, F.~C. Seman, M.~I. Abbasi, N.~F. Sallehuddin,
  A.~Y. I.~Ashyap, and M.~R. Kamarudin, ``Aspects of efficiency enhancement in
  reflectarrays with analytical investigation and accurate measurement,''
  \emph{Electron.}, vol.~9, no.~11, 2020.

\bibitem{insertion_loss_table}
B.~Cetindogan, B.~Ustundag, E.~Turkmen, M.~Wietstruck, M.~Kaynak, and
  Y.~Gurbuz, ``A {D}-band {SPDT} switch utilizing reverse-saturated {SiGe HBTs}
  for dicke-radiometers,'' in \emph{Proc. IEEE GeMiC}, Freiburg, Germany, 2018,
  pp. 47--50.

\bibitem{8874296}
O.~Cojocari, D.~Moro-Melgar, and I.~Oprea, ``High-power mm-wave frequency
  multipliers,'' in \emph{Proc. IRMMW-THz}, Paris, France, 2019.

\bibitem{SUYAMA20212020FGI0002}
S.~Suyama, T.~Okuyama, Y.~Kishiyama, S.~Nagata, and T.~Asai, ``A study on
  extreme wideband {6G} radio access technologies for achieving 100{Gbps} data
  rate in higher frequency bands,'' \emph{IEICE Trans. Commun.}, vol. E104.B,
  no.~9, pp. 992--999, 2021.

\bibitem{Terraameta_EuCAP2024}
G.~C. Alexandropoulos, A.~Clemente, S.~Matos, R.~Husbands, S.~Ahearne, Q.~Luo,
  V.~Lain-Rubio, T.~Kürner, and L.~M. Pessoa, ``Reconfigurable intelligent
  surfaces for {THz}: Signal processing and hardware design challenges,'' in
  \emph{Proc. EuCAP}, Glasgow, Scotland, 2024.

\bibitem{RIS_rich_scattering}
G.~C. Alexandropoulos, N.~Shlezinger, and P.~del Hougne, ``Reconfigurable
  intelligent surfaces for rich scattering wireless communications: {R}ecent
  experiments, challenges, and opportunities,'' \emph{IEEE Commun. Mag.},
  vol.~59, no.~6, pp. 28--34, 2021.

\bibitem{Transmitarray_THz_2023}
O.~Koutsos, F.~F. Manzillo, M.~Caillet, R.~Sauleau, and A.~Clemente,
  ``Experimental demonstration of a $43$-{dBi} gain transmitarray in {PCB}
  technology for backhauling in the $300$-{GHz} band,'' \emph{IEEE Trans.
  Terahertz Sci. Technol.}, vol.~13, no.~5, pp. 485--492, 2023.

\bibitem{Eucap2024RISfab}
X.~Ma, Y.~Zhou, Q.~Luo, Y.~Ma, K.~Stylianopoulos, and G.~C. Alexandropoulos,
  ``$1$-bit {subTHz RIS} with planar tightly coupled dipoles: {B}eam shaping
  and prototypes,'' in \emph{Proc. EuCAP}, Glasgow, Scotland, 2024.

\bibitem{Sobo22}
J.~Sobolewski and Y.~Yashchyshyn, ``State of the art sub-terahertz switching
  solutions,'' \emph{IEEE Access}, vol.~10, pp. 12\,983--12\,999, 2022.

\bibitem{Liu22}
R.~Liu, J.~Dou, P.~Li, J.~Wu, and Y.~Cui, ``Simulation and field trial results
  of reconfigurable intelligent surfaces in {5G} networks,'' \emph{{IEEE}
  Access}, vol.~10, pp. 122\,786--122\,795, 2022.

\bibitem{Mansur24}
T.~Singh, N.~K. Khaira, M.~Repeta, and R.~R. Mansour, ``Phase-change {RF}
  devices for future communications: Phase-change materials and devices for
  reconfigurable {RF} front-ends: State-of-the-art and future perspectives,''
  \emph{IEEE Microw. Mag.}, vol.~25, no.~2, pp. 18--38, 2024.

\bibitem{Wain21}
N.~Wainstein, G.~Adam, E.~Yalon, and S.~Kvatinsky, ``Radiofrequency switches
  based on emerging resistive memory technologies - {A} survey,'' \emph{Proc.
  {IEEE}}, vol. 109, no.~1, pp. 77--95, 2021.

\bibitem{Rebeiz12}
M.~Uzunkol and G.~M. Rebeiz, ``140–220 {GHz} {SPST} and {SPDT} switches in 45
  nm {CMOS} {SOI},'' \emph{IEEE Microw. Wireless Compon. Lett.}, vol.~22,
  no.~8, p. 412–414, 2012.

\bibitem{Seo23}
W.~Seo, S.~Kim, B.~Ko, H.~Jhon, and J.~Kim, ``High-powered {RF SOI} switch with
  fast switching time for {TDD} mobile applications,'' \emph{IEEE Access},
  vol.~11, p. 7277–7282, 2023.

\bibitem{Cress14}
P.~Song, R.~L. Schmid, A.~C. Ulusoy, and J.~D. Cressler, ``A high-power,
  low-loss {W}-band {SPDT} switch using {SiGe} {PIN} diodes,'' in \emph{Proc.
  IEEE RFIC}, Tampa, USA, Jun. 2014.

\bibitem{gemic18}
B.~Cetindogan, B.~Ustundag, E.~Turkmen, M.~Wietstruck, M.~Kaynak, and
  Y.~Gurbuz, ``A {D}-band {SPDT} switch utilizing reverse-saturated {SiGe}
  {HBTs} for dicke-radiometers,'' in \emph{Proc. IEEE GeMiC}, Freiburg,
  Germany, 2018, pp. 47--50.

\bibitem{Intel20}
H.~W. Then \emph{et~al.}, ``Advances in research on 300mm {Gallium}
  {Nitride}-on-{Si}(111) {NMOS} transistor and silicon {CMOS} integration,'' in
  \emph{Proc. IEEE IEDM}, San Francisco, USA, 2020.

\bibitem{Longhi24}
P.~E. Longhi, L.~Pace, F.~Costanzo, W.~Ciccognani, S.~Colangeli, R.~Giofrè,
  R.~Leblanc, A.~Suriani, F.~Vitobello, and E.~Limiti, ``32–36-{GHz}
  single-chip front-end {MMIC} featuring 35-{dBm} output power and 3.2-{dB}
  noise figure with 60- and 100-nm {GaN/Si HEMTs},'' \emph{IEEE Trans. Microw.
  Theory Tech.}, vol.~72, no.~1, pp. 160--172, 2024.

\bibitem{reichel2018electrically}
K.~S. Reichel, N.~Lozada-Smith, I.~D. Joshipura, J.~Ma, R.~Shrestha, R.~Mendis,
  M.~D. Dickey, and D.~M. Mittleman, ``Electrically reconfigurable terahertz
  signal processing devices using liquid metal components,'' \emph{Nature
  Commun.}, vol.~9, no.~1, 2018.

\bibitem{jackel1982electrowetting}
J.~Jackel, S.~Hackwood, and G.~Beni, ``Electrowetting optical switch,''
  \emph{Applied Physics Letters}, vol.~40, no.~1, pp. 4--5, 1982.

\bibitem{Ambacher19}
F.~Thome, P.~Bruckner, R.~Quay, and O.~Ambacher, ``Millimeter-wave single-pole
  double-throw switches based on a 100-nm gate-length {AlGaN/GaN-HEMT}
  technology,'' in \emph{Proc. IEEE MTT-S IMS}, Boston, USA, 2019.

\bibitem{Kiazadeh2023}
A.~Kiazadeh, J.~Deuermeier, E.~Carlos, R.~Martins, S.~Matos, F.~M. Cardoso, and
  L.~M. Pessoa, ``Concept paper on novel radio frequency resistive switches,''
  in \emph{Proc. ACM Int. Symp. Nanosc. Architect.}, New York, USA, 2023.

\bibitem{sen2008microscale}
P.~Sen and C.-J.~C. Kim, ``Microscale liquid-metal switches—{A} review,''
  \emph{IEEE Trans. Ind. Electron.}, vol.~56, no.~4, pp. 1314--1330, 2008.

\end{thebibliography}
\end{document}